# Organized Complexity:
# Is Big History a Big Computation?

Jean-Paul Delahaye
Centre de Recherche en Informatique, Signal et
Automatique, UMR CNRS 9189
Université de Lille 1,
jean-paul.delahaye@univ-lille1.fr

Clément Vidal
Center Leo Apostel &
Evolution Complexity and Cognition,
Vrije Universiteit Brussel,
contact@clemvidal.com

**Abstract:** The concept of "logical depth" introduced by Charles H. Bennett (1988) seems to capture, at least partially, the notion of *organized complexity*, so central in big history. More precisely, the increase in organized complexity refers here to the wealth, variety and intricacy of structures, and should not be confused with the increase of *random complexity*, formalized by Kolmogorov (1965). If Bennett is right in proposing to assimilate organized complexity with "computational content", then the fundamental cause of the increase of complexity in the universe is the existence of computing mechanisms with memory, and able to cumulatively create and preserve computational contents. In this view, the universe computes, remembers its calculations, and reuses them to conduct further computations. Evolutionary mechanisms are such forms of cumulative computation with memory and we owe them the organized complexity of life. Language, writing, culture, science and technology can also be analyzed as computation mechanisms generating, preserving and accelerating the increase in organized complexity. The main unifying theme for big history is the energy rate density, a metric based on thermodynamics. However useful, this metric does not provide much insight into the role that information and computation play in our universe. The concept of "logical depth" provides a new lens to examine the increase of organized complexity. We argue in this paper that organized complexity is a valid and useful way to make sense of big history. Additionally, logical depth has a rigorous formal definition in theoretical computer science that hints at a broader research program to quantify complexity in the universe.

**Keywords**: organized complexity, Kolmogorov complexity, logical depth, big history, cosmic evolution, evolution, complexity, complexification, computation, artificial life, philosophy of information

## 1  Introduction

The core concept of big history is the increase of complexity (Christian 2004). Currently, it is mainly explained and analyzed within a thermodynamic framework, with the concept of energy rate density (Chaisson 2001; 2011).

However, even if energy is universal, it doesn't capture informational and computational dynamics, central in biology, language, writing, culture, science and technology. Energy is, by definition, not an informational concept. Energy can produce poor or rich interactions; it can be wasted or used with care. The production of computation by unit of energy varies sharply, from device to device. For example, a compact disc player produces much less computation per unit of energy than a regular laptop. Furthermore, Moore's law shows that from computer to computer, the energy use per computation decreases quickly with each new generation of microprocessor.

Since the emergence of life, living systems have evolved memory mechanisms (RNA, DNA, neurons, culture, technologies) storing information about complex structures. In that way, evolution needs not to start from scratch, but can build on previously memorized structures. Evolution is thus a cumulative process based on useful information, not on energy, in the sense that energy is necessary, but not sufficient. Informational and computational metrics are needed to measure and understand such mechanisms.

We take a computational view on nature, in the tradition of digital philosophy (e.g. Zuse 1970; Chaitin 2006; Lloyd 2005; Wolfram 2002; Floridi 2003). In this framework, cosmic evolution is essentially driven by memory mechanisms that store previous computational contents, on which further complexity can be built.

We first give a short history of information theories, starting with Shannon, but focusing on algorithmic information theory, which goes much further. We then elaborate on the distinction between *random complexity*, formalized by Kolmogorov (1965), and *organized complexity*, formalized by Bennett (1988). Kolmogorov complexity (K) is a way to measure *random complexity*, or the *informational content* of a string. It is defined as the size of the shortest program producing such a string.

This tool has given rise to many applications, such as automatic classification in linguistics (Cilibrasi and Vitanyi 2005; Li et al. 2004), automatic generation of phylogenetic trees (Varré, Delahaye, and Rivals 1999), or to detect spam (Belabbes and Richard 2008).

Bennett's logical depth does not measure an informational content, but a *computational content*. It measures the time needed to compute a certain string S from a short program. A short program is considered as a more probable origin of S than a long program. Because of this central inclusion of time, a high (or *deep*) value in logical depth means that the object has had a rich causal history. In this sense, it can be seen as a mathematical and computational formalization of the concept of history. More broadly construed (i.e. not within the strict formal definition), we want to show that modern informational, computational and algorithmic theories can be used as a conceptual toolbox to analyze, understand and explore the rise of complexity in big history.

We outline a research program based on the idea that what reflects the increase of complexity in cosmic evolution is the computational content, that we propose to assimilate with logical depth, i.e. the associated mathematical concept proposed by Bennett. We discuss this idea at different levels, formally, quasi-physically and philosophically. We end the paper with a discussion of issues related to this research program.

# 2 A very short history of information theories

## 2.1 Shannon information theory

The Shannon entropy (Shannon 1948) of a sequence S of *n* characters is a measure of the information content of S when we suppose that every character C has a fixed probability *pr*(C) to be in position *i* (the same for every position). That is:

$$H(S) = n[-\sum_{C} pr(C)(log_2(pr(C)))]$$

If we know only this probabilistic information about S, it is not possible to compress the sequence S in another sequence of bits of length less then H(S). Actual compression algorithms applied to texts do search and use many other regularities beyond the relative frequency of letters. This is why Shannon entropy does not give the real minimal length in bits of a possible compressed version of S. This minimal length is given by the Kolmogorov complexity of S that we will now introduce.

## 2.2 Algorithmic information theory

Since 1965, we've seen a renewal of informational and computational concepts, well beyond Shannon's information theory. Ray Solomonoff, Andreï Kolmogorov (1965), Leonid Levin, Pier Martin-Löf (1966), Gregory Chaitin, Charles Bennett are the first contributors of this new science (see Li and Vitányi 2008 for details), which is based on the mathematical theory of computability born with Alan Turing in the 1930s.

The Kolmogorov complexity K(S) of a string S is the length of the smallest program S* written in binary code and for a universal computer that produces S. This is the absolute informational content or incompressible information content of S, or the algorithmic entropy of S.

Kolmogorov complexity is also called interchangeably *informational content* or *incompressible informational content* or *algorithmic entropy* or *Kolmogorov-Chaitin algorithmic complexity* or *program-size complexity*.

The invariance theorem states that K(S) does not really depend on the used programming language, provided the language is universal (capable to define every computable function).

The Kolmogorov complexity is maximal for random sequences: a random sequence cannot be compressed. This is why K(S) is sometimes called *random complexity* of S.

## 2.3 Logical depth - Computational content

Kolmogorov complexity is an interesting and useful concept, but it is an error to believe that it measures the value of the information contained in S. Not all information is useful: for example, the information in a sequence of heads and tails generated by throwing a coin is totally useless. Indeed, if a program needs to use a random string, another random string would also do the job, which means that the particular random string chosen is not important. Kolmogorov complexity is a useful notion for defining the absolute notion of a random sequence (Martin-Löf 1966), but it is not capturing the notion of organized complexity.

Charles H. Bennett has introduced another notion, the "logical depth of S". It tries to measure the real value of the information contained in S, or as he proposed its "computational content" (to be opposed to its 'informational content"). A first attempt to formulate Bennett's idea is to say that the logical depth of S, LD(S) is the time it takes for the shortest program of S, S*, to produce S. A more detailed study and discussion about the formulation can be found in (Bennett 1988).

Various arguments have been formulated that make plausible that indeed the logical depth of Bennett, LD(S), is a measure of the computational content of S, or of the quantity of non trivial structures in S. To contrast it to "random complexity", we say that it is a measure of "organized complexity".

An important property of LD(S) is the *slow growth's law* (see Bennett 1988): an evolutionary system S($t$) cannot have its logical depth LD(S($t$)) that grows suddenly.
This property (which is not true for the Kolmogorov complexity) seems to correspond to the intuitive idea that in an evolutionary process, whether it is biological, cultural or technological, the creation of new innovative structures cannot be quick.

Variants of logical depth have been explored (Lathrop and Lutz 1999; Antunes et al. 2006; Doty and Moser 2007), as well as other similar ideas, such as *sophistication (Koppel 1987; 1995; Koppel and Atlan 1991; Antunes and Fortnow 2003)*, *facticity (Adriaans 2009; 2012)* or *effective complexity (Gell-Mann and Lloyd 1996; 2004)*. Studies have established properties of these measures, and have discussed them (Antunes, Souto, and Teixeira 2012; Bloem, Rooij, and Adriaans 2015). Importantly, results show that these various notions are closely related (Ay, Muller, and Szkola 2010; Antunes et al. 2016). In this paper we focus on logical depth, whose definition is general, simple and easy to understand.

## 3 Outline of a research program

### 3.1 Three levels of analysis

Let us first distinguish three conceptual levels of the notion of computational content: *mathematical*, *quasi-physical* and *philosophical*.

First, we presented the notion of computational content as the logical depth, as defined by Bennett. Other formal definitions of computational content may be possible, but this one has proven to be robust. This definition has been applied to derive a method to classify and characterize the complexity of various kinds of images (Zenil, Delahaye, and Gaucherel 2012). More applications promise to be successful, in the same way as Kolmogorov complexity proved useful.

Second, we have the quasi-physical level, linking computation theory with physics (Bennett 2012; Feynman 1998). This has not yet been developed in a satisfactory manner. Maybe this would require physics to consider a fundamental notion of computation, in the same way as it integrated the notion of information (used for example in thermodynamics). The transfer of purely mathematical or computer science concepts into physics is a delicate step. Issues relate for example to the thermodynamics of computation, the granularity of computation we look at, or the design of hardware architectures actually possible physically.

The concept of thermodynamic depth introduced by Seth Lloyd et Heinz Pagels (1988) is defined as "the amount of entropy produced during a state's actual evolution". It is a first attempt to translate Bennett's idea in a more physical context. However the definition is rather imprecise and it seems not really possible to use it in practice. It is not even clear that it reflects really the most important features of the mathematical concept, since "thermodynamical depth can be very system dependant: some systems arrive at a very trivial state through much dissipation; others at very non trivial states with little dissipation" (Bennett 1990, 142).

Third, the philosophical level brings the bigger picture. It captures the idea that building complexity takes time and interactions (computation time). Objects measured with a deep computational content necessarily have a rich causal history. It thus reflects a kind of historical complexity. Researchers in various fields have already recognized its use (Gell-Mann 1994; Danchin 2003; Mitchell 2009; Mayfield 2013; Steinhart 2014; Dessalles, Gaucherel, and Gouyon 2016).

This philosophical level may also hint at a theory of value based on computational content (Steinhart 2014, chap. 73; Delahaye and Vidal 2018). For example, a library has a huge computational content, because it is the result of many brains who worked to write books. Burning a library can thus be said to be unethical.

## 3.2 Computer simulations

A major development of modern science is the use of computer simulations. Simulations are essential tools to explore dynamical and complex interactions that cannot be explored with simple equations. Since the most important and interesting scientific issues are complex, simulations will likely be used more and more systematically in science (Vidal 2008).

The difficulty with simulations is often to interpret the results. We propose that Kolmogorov complexity (K) and logical depth (LD) would be valuable tools to test various hypotheses relative to the growth of complexity. Approximations of K and LD have already been applied to classify the complexity of animal behavior. These algorithmic methods do validate experimental results obtained with traditional cognitive-behavioural methods (Zenil, Marshall, and Tegnér 2015).

For an application of K-complexity and LD to an artificial life simulation, see for example the work of Gaucherel (2014), comparing a Lamarkian algorithm with a Darwinian algorithm in an artificial life simulation. Gaucherel proposes the following three-step methodology:

> (1) identification of the shortest program able to numerically model the studied system (also called the Kolmogorov–Solomonoff complexity); (2) running the program, once if there are no stochastic components in the system, several times if stochastic components are there; and (3) computing the time needed to generate the system with LD complexity.

More generally, in the domain of Artificial Life, it is fundamental to have metric monitoring if the complexity of the simulated environment really increases. Testing the logical depth of entities in virtual environments would prove very useful.

### 3.3 Emergy and logical depth

In systems ecology, an energetic counterpart to the notion of computational content has been proposed. It is called *emergy* (with an "m") and is defined as the value of a system, be it living, social or technological, as measured by the solar energy that was used to make it (e.g. Odum 2007). This is very similar to the logical depth, defined by the quantity of computation that needs to be performed to make a structured object.

Does this mean that energetic content (emergy) and computational content are one and the same thing? No, and one argument amongst many others is that the energetic content to produce a computation diminishes tremendously with new generations of computers (c.f. Moore's law).

## 4 Discussion

We formulate here a few questions that the reader may have, and propose some answers.

**Before the emergence of life, does cosmic evolution produces any computational content?**

Yes, but the memorization of calculus is non-existent or very limited. A computation does not necessarily mean a computation with memorization. For example, atoms such as H or molecules such as $H_2O$ are all the same, there is no memory of what has happened to a particular atom or molecule. What lacks in these cases is computation with a memory mechanism.

The increase of complexity accelerates with the emergence of more and more sophisticated and reliable memory mechanisms. In this computational view, the main cosmic evolution threshold is the emergence of life, because it creates a memory mechanism in the universe (RNA/DNA). From a cosmic perspective, complexity transitions have decelerated from the Big Bang to the origin of life, and started to accelerate since life appeared (Aunger 2007). The emergence of life thus constitutes the tipping point in the dynamics of complexity transitions.

Furthermore, evolutionary transitions are marked with progress in the machinery to manipulate information, particularly regarding the *memorization* of information. For example, we can think of RNA/DNA, nervous systems, language, writing, and computers as successive revolutions in information processing (Dawkins 1995).

**Why would evolution care about minimal-sized programs?**

We care about *short* programs, not necessarily minimally-sized programs proven to be so. The shortest program (or a near shortest program) producing S is the most probable origin for S. Let us illustrate this point with a short story. Imagine that you walk in the forest, and find engraved on a tree trunk 1,000,000 digits of π, written in binary code. What is the most probable explanation of this phenomenon? There are $2^{1\,000\,000}$ strings of the same size, so the chance explanation has to be excluded. The first plausible explanation is rather that it is a hoax. Somebody computed digits of π, and engraved them here. If a human did not do it, a physical mechanism may have done it, that we can equate with a short program producing π. The likely origin of the digits of π is a short program producing them, not a long program of the kind print(S), which would have a length of about one million.

Another example from the history of science is the now refuted idea of spontaneous generation (Strick 2000). From our computational perspective, it would be extremely improbable that sophisticated and complex living systems would appear in a few days. The slow growth law says that they necessarily needed time to appear.

**Couldn't you have a short program computing for a long time, with a trivial output, which would mean that a trivial structure would have a deep logical depth?**

Of course, programs computing a long time and producing a trivial output are easy to write. For example, it is easy to write a short program, computing for a long time, and producing a sequence of 1000 zeros. This long computation wouldn't give the logical depth the string, because there also a shorter program computing much more rapidly and producing these 1000 zeros. This means that objects with a deep logical depth can't be trivial.

**Why focus on decompression times and not compression times?**

The compression time is the time necessary to resolve a problem: knowing S, find the shortest (or a near shortest) program producing S.

By contrast, the decompression time is the time necessary to produce the sequence S from a near shortest program that produces S. It is thus a very different problem from compression.

If we imagine that the world contains many explicit or implicit programs —and we certainly can think our world as a big set of programs producing objects— then the probability of an encounter with a sequence S depends only on the time necessary for a short program to produce S (at first glance, only short program exist).

**Complexity should be defined dynamically, not statically.**

A measure is by definition something static, at one point in time. However, we can compare two points in time, and thus study the relative LD, and the dynamics of organized complexity.

Let us take a concrete example. What is the difference in LD-complexity between a living and a dead body? At the time of death, the computational content would be almost the same for both. This is because the computational content measures the causal history. A dead person still has had a complex history. Other metrics may be used to capture more dynamical aspects such as informational flows or energy flows.

# 5 Conclusion

To sum up, we want to emphasize again that random complexity and organized complexity are two distinct concepts. Both have strong theoretical foundations, and have been applied to measure the complexity of particular strings. More generally, they can be applied in practice to assess the complexity of some computer simulations. In principle, they may thus be applied to any physical object, given that it is modelled digitally or in a computer simulation.

Applied to big history, organized complexity suggests that evolution retains computational contents via memory mechanisms, whether they are biological, cultural or technological. Organized complexity further indicates that major evolutionary transitions are linked with the emergence of new mechanisms that compute and memorize.

Somewhat ironically, complexity measures in big history have neglected history. We have argued that the computational content, reflecting the causal history of an object and formalized as logical depth − as defined by Bennett − is a promising complexity metric in addition to existing energetic metrics. It may well become a general measure of complexity.